\RequirePackage{fix-cm}
\documentclass[smallextended]{svjour3}       
\smartqed  
\usepackage{graphicx}
\usepackage{diagbox}
\usepackage[authoryear,longnamesfirst]{natbib}
\usepackage{wrapfig}
\usepackage[ruled,vlined]{algorithm2e}
\usepackage{algpseudocode} 
\usepackage{multirow}
\usepackage{adjustbox}

\graphicspath{ {Figures/} }

\journalname{Education and Information Technologies}
\begin{document}

\title{A weighted unified informetrics based on Scopus and WoS}


\author{Parul Khurana         \and
        Geetha Ganesan         \and
        Gulshan Kumar         \and
        Kiran Sharma 
}


\institute{Parul Khurana \at
              School of Computer Applications, Lovely Professional University, Phagwara, Punjab-144401, India \\
              \email{parul.khurana@lpu.co.in}           
           \and
           Geetha Ganesan \at
              Advanced Computing Research Society, Porur, Chennai-600116, India \\
              \email{gitaskumar@yahoo.com}           
                         \and
           Gulshan Kumar \at
              School of Computer Science and Engineering, Lovely Professional University, Phagwara, Punjab-144401, India \\
              \email{gulshan3971@gmail.com}           
                         \and
           Kiran Sharma \at
              School of Engineering and Technology, BML Munjal University, Gurugram, Haryana-122413, India \\
              \email{kiran.sharma@bmu.edu.in}           
}

\date{Received: date / Accepted: date}

\maketitle

\begin{abstract}
Numerous indexing databases keep track of the number of publications, citations, etc. in order to maintain the progress of science and individual.  However, the choice of journals and articles varies among these indexing databases, hence the number of citations and $h$-index varies. There is no common platform exists that can provide a single count for the number of publications, citations, $h$-index, etc.  To overcome this limitation, we have proposed a weighted unified informetrics, named ``conflate''.  The proposed system takes into account the input from multiple indexing databases and generates a single output.  Here,  we have used the data from Scopus and WoS to generate a conflate dataset. Further, a comparative analysis of conflate has been performed with Scopus and WoS at three levels: author, organization, and journal.  Finally,  a mapping is proposed between research publications and distributed ledger technology in order to provide a transparent and distributed view to its stakeholders.
\end{abstract}

\section{Introduction}

Bibliometrics refers to the use of quantitative measures and statistical methods to analyze the impact of scientific contributions \citep{umate41bibliometric}. It helps in describing publication patterns, measuring the research quality and impact, and determining the influence of authors with different authors involved in publications \citep{daim2006forecasting, archambault2009comparing, albort2018assessing}. Scientific contributions work as a driving force for the continuous growth of science and society \citep{van2003use, janowicz2018prospects}. These contributions receive citations which are considered as an important indicator of influence in the research publications process and are explored and studied widely in this digital world \citep{pringle2008trends, sharma2020growth}. Citations provide a quantitative evaluation to measure research outputs, utilities, emergence, and extent of scientific communication to all the interested parties  \citep{meho2009assessing, hoffman2019scholarly}. Citation data is often impacted by the coverage of specific bibliographic databases because they collect the citations received by the publications indexed by them \citep{bar2007some}. Such bibliographic databases offer scientific activities, number of publications, number of citations, and quality, importance, and impact of publications and their peer evaluation for bibliometric analysis \citep{martin1996use, abramo2011evaluating}.

Bibliographic databases like Elsevier's Scopus and Thomson Reuter's WoS are being used worldwide to produce comparative statistics for their bibliometricians. Both are working continuously in this dynamic field so that required informetrics may be empowered with their databases \citep{bakkalbasi2006three, neuhaus2008data, adriaanse2011comparing, franceschini2016empirical}. Comparative statistics provided by bibliographic databases are used by funding agencies, government bodies, promotion committees, and other stakeholders to measure the quality and impact of authors \citep{meho2006new, martin2018google}. Hence, bibliometric analysis has emerged as a powerful tool and partial system for ranking and accreditation agencies as well \citep{alvarez2018resources, oliveira2018prospective}.

Different authors have shown the comparison of Scopus and WoS in various literatures for the coverage of journals \citep{mongeon2016journal, liu2021same}, quality of content  \citep{vafaeian2011comparative}, bibliometric statistics \citep{de2012coverage, bartol2015bibliometric}, subject categorization \citep{powell2017coverage}, features and capabilities \citep{thelwall2018dimensions, martin2019google}, citations counts, tracking and indexing \citep{jacso2005we, vaughan2008new, li2010citation}, author, university and country rankings \citep{lopez2009comparing}, language coverage \citep{bartol2014assessment}, longitudinal and cross-disciplinary comparisons \citep{norris2007comparing}, content comprehensiveness and searching capabilities \citep{burnham2006scopus, aghaei2013comparison}. The prolific growth of both bibliographic databases have created new opportunities in terms of publications, citations and bibliometric indicators \citep{adriaanse2013web} as well.

Due to the availability of the number of bibliographic databases, interviewers and hiring agencies ask authors to provide publications, citations, and $h$ index count - bibliographic databases wise. This situation in India has raised a requirement of unified informetrics where an author can provide a single count for his/her bibliometric statistics across different bibliographic databases. Hence, we conducted this study to calculate a weighted unified informetrics at three levels such as author, organization, and journal. We have used bibliographic databases such as Scopus and WoS due to their indexing age, availability of data, and authenticity. The objectives of our study are:

\begin{enumerate}
\item To propose a common platform that can provide a single article count, citation count, and $h$-index in the education field.

\item To check the statistical validity of the proposed platform in terms of the number of articles,  the number of citations, and $h$-index at author, organization, and journal level.

\item Technological enhancement of the proposed platform by use of distributed ledger technology.
\end{enumerate}

Further, the study is organized as follows: Section~\ref{sec:data} describes data and methods including data selection, data filtration, data extraction, data analysis, and concept of pay-off matrix. Results are explained in Section~\ref{sec:stats}. Section~\ref{sec:tech} describes the technological enhancement with the concept of distributed ledger technology. Finally, summary is given in Section~\ref{sec:conclusion}.

\section{Data and methods}
\label{sec:data}
\subsection{Data selection and filtration}

Data selection is performed at three levels:
\begin{itemize}
\item \textit{Author level:} 400 faculty profiles out of 6316 profiles from various disciplines are accessed from ``Monash University'', a public university in Melbourne, Australia \citep{monash2020}. The choice of the data is mainly due to the openly available information of faculties especially ORCID ID, Scopus ID, and WoS ID. All 400 selected faculty profiles have an ORCID ID, Scopus ID, and WoS ID.

\item \textit{Organization level:} Top 100 Indian institutes out of 200 ranked institutions based on \textit{National Institutional Ranking Framework} (NIRF) \citep{nirf2020} are accessed. 

\item \textit{Journal level:} A random selection of 1000 journals listed in both Scopus and WoS is used.
\end{itemize}

\begin{figure}[h!]
    \centering
    \includegraphics[width=\linewidth]{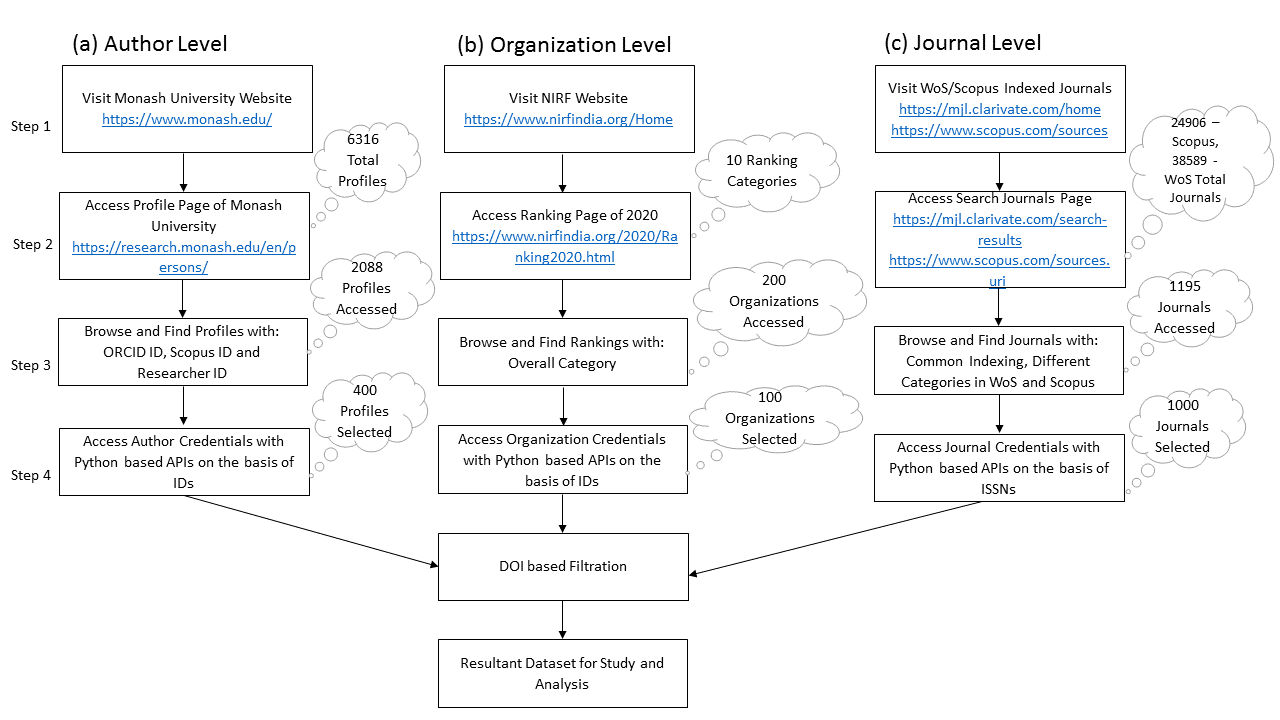}
    \caption{Flowchart demonstrate the process of visiting the author's, organization's and journal's profile.}
    \label{fig1}
\end{figure}

Data filtration is performed on the basis of DOI (Digital Object Identifier). It provides a unique authentication to the publications \citep{simmonds1999digital, chandrakar2006digital}. It’s desirable that articles carrying DOI numbers must be considered only for any kind of evaluation and calculation to establish scientific assignments \citep{gorraiz2016availability}. Hence, while combining articles across multiple databases, we have considered the articles with DOI numbers only (see flowchart in Figure~\ref{fig1}). To perform this study, data from Scopus and WoS was obtained with Python-based APIs \citep{pybliometrics, wos}. The primary reason for data selection from Scopus and WoS is arbitrary and the availability of data.

\subsection{Data extraction and analysis}

For data extraction from both Scopus and WoS, we require three inputs at three levels. An ORCID ID is required for authors’ information, organization name for university/institute access, and ISSN for journal information. 
For a given author, based on his/her ORCID ID, we retrieved the number of publications from both Scopus and WoS.  Further, we filtered those publications based on the DOI and got the required data set for the given author for both Scopus and WoS.  The next step is to look after the number of citations received by the given author for each publication. We examined each and every citation received on all publications for a given author and filtered only those citations that have DOI associated to them. The existence of DOI is necessary to condition in order to match the given record among multiple databases.  The collection of similar records and respective citations based on DOI for a given author among Scopus and WoS will lead to a new filtered database named ``conflate''.
Algorithm~\ref{Algo1} describes the steps for extracting article and citation details for the given ORCID ID. Similarly, the process is repeated at the organization and the journal levels.
The conflate database of citations generated in algorithm~\ref{Algo1} is used further for the analysis. First,  for a given author, we filtered the  \textit{common} and the \textit{unique} citations (publications records) from Scopus and WoS. Then,  a weight is assigned to both common and unique citations (see section~\ref{sec:pay_off}). These new weighted citations will be our conflate data for a given author. Algorithm~\ref{Algo2} describes the process in sequence.

\SetKwInput{KwInput}{Input}                
\SetKwInput{KwOutput}{Output}              
\begin{algorithm}[H]
\caption{Generation of conflate based citations database} \label{Algo1}
\DontPrintSemicolon
  
  \KwInput{ORCID ID, ISSN, or Organization Name}
  \KwOutput{Conflated citations from $N$ databases}
  \tcc{Check number of articles in $N$ databases}
\nl \While{$N>0$} 
  {
        $A_1, A_2, ..., A_N \in DB_1, DB_2, ..., DB_N$, $ \forall  \textrm{\textit{doi} exists}$\\
        where $A_1$ represents articles \textit{doi} list for $DB_1$
  }
 \tcc{Find \textit{doi} of each citation}
\nl  \For{each \textit{doi} in $[A_1, A_2, ..., A_N]$}
  {
    $N_c$ := number of citations\\
    \For{each $N_c$}
    {
        \If{doi exists}
        {$CD_1$ := doi}  \tcp*{CD is citer's doi}
        \Else
        {skip}
    }
  }
\nl Repeat step 2 for all  $A_1,A_2,...,A_N$ and get  $CD_1,CD_2,...,CD_N$\\
\nl $CD$ := $CD_1 \cup CD_2 \cup ... \cup CD_N$\\
where $CD$ contains only those citations for a given ORCID ID whose \textit{doi's} exists (including common \textit{doi's} in all DB's)

\end{algorithm}

\SetKwInput{KwInput}{Input}                
\SetKwInput{KwOutput}{Output}              
\begin{algorithm}[H]
\caption{Generation of $h$ index and related indicators } \label{Algo2}
\DontPrintSemicolon
  
  \KwInput{Conflate based citations from $N$ databases for a given ORCID ID, ISSN, or Organization Name}
  \KwOutput{Updated $h$ index and related indicators} 
  
  \KwData{List of citer's \textit{doi's}, $CD_N$, $A_N$}  \tcp*{Computed in Algorithm~\ref{Algo1}}
 
\nl common := $CD_1 \cap CD_2 \cap ... \cap CD_N$\\ \tcp*{get the common doi's in all DB}
\nl unique := $CD-common$\\ \tcp*{only in individual DB's}

\nl Compute weightage of all \textit{doi's}\\
     \tcp*{Give full weightage, i.e.1 to doi's in \textit{common} set (in step 1)}
     S1 = Number of \textit{doi's} in \textit{common}\\
     \tcp*{Give $\frac{1}{N}$ weightage to doi's in \textit{unique} set (in step 2)}
    S2 = Number of \textit{doi's} in \textit{unique}*$ \frac{1}{N}$\\
   \tcp*{Final weighted citation}
    $S = ceil(S1+S2)$\\

  \tcc{Find unique \textit{doi's} in CD}
\nl $C= unique(CD)$\\
\nl Compute $h$-index and related indicators on $C$\\


\end{algorithm}

\subsection{Pay off matrix}
\label{sec:pay_off}

For ranking the universities, various agencies like NIRF and Times Higher Education (THE) use databases of their choice like ~\cite{THE2020} use Scopus database and NIRF gives equal weightage to all databases~\citep{nirf2020}.  As all bibliographic databases carry their own special features and fields of information, we cannot decide that one indexing database is better than others~\citep{papic2017informetrics, rose2019pybliometrics, yelne2021assessment, khurana2021impact}. Hence,  it is not appropriate to give more weightage to any specific database over others. However, a weighted measure can be used on such databases.  On the basis of the above-discussed limitation, we have proposed the weight assignment as given in Table~\ref{Table:1}.

For a given publication (DOI exists in both Scopus and WoS),  if a citation with DOI exists in both Scopus and WoS, then it would be considered as a \textit{common} citation and would be assigned a weight 1. On the other hand, a \textit{unique} citation, i.e. either in Scopus or WoS, the assigned weight would be 0.5 (in the case of $N$ databases, the pay off will be $1/N$). Table~\ref{Table:2} demonstrates three examples with different scenarios.
\begin{itemize}
\item  Example 1: Publication ($P$) has received 3 citations in Scopus and 0 in WoS. Hence, there is no \textit{common} citation and we have left with 3 \textit{unique} citations. So the proposed number of citations of $P$ would be 2.

\item  Example 2: $P$ has received 3 citations in Scopus and 3 in WoS.  Let's say there are 2 \textit{common} citations and 2 \textit{unique} citations (1 from each database). So the proposed number of citations of $P$ would be 3.
\item Example 3:  When all citations are \textit{common} citations. No payoff is assigned.
\end{itemize}

\begin{table}[h!]
\centering
\caption{The pay off matrix demonstrates the weight for \textit{common} and \textit{unique} citations.  Three possible cases are: (i) both databases have a common citation,  (ii) either or both databases have unique citation,  and (iii) no citation.}
\begin{tabular}{|c|c|c|}
\hline
\diagbox{Scopus}{WoS}                       & Received citations & Not received citations \\ \hline
Received citations     & (0.5, 0.5)         & (0.5, 0)               \\ \hline
Not received citations & (0, 0.5)           & (0, 0)                 \\ \hline
\end{tabular}
\label{Table:1}
\end{table}

\begin{table}[h!]
\centering
\caption{Weight assignment to a publication $P$ for $N=2$ databases, Scopus (S) and WoS (W). Weight assigned to citations as \textit{common}=1 and \textit{unique}=0.5. }
\begin{tabular}{|c|c|c|c|c|c|c|}
\hline
\multirow{3}{*}{Example} & \multicolumn{5}{c|}{Citations}                                      & \multirow{3}{*}{\begin{tabular}[c]{@{}c@{}}Weight = Common\\ + Unique/N\end{tabular}} \\ \cline{2-6}
                         & \multicolumn{2}{c|}{Total} & Common   & \multicolumn{2}{c|}{Unique} &                                                                                       \\ \cline{2-6}
                         & S            & W           & S+W      & S            & W            &                                                                                       \\ \hline
1                        & 3            & 0           & 0        & 3            & 0            & \begin{tabular}[c]{@{}c@{}}0 + 3/N = 1.5 $\approx 2$\end{tabular}               \\ \hline
2                        & 3            & 3           & 2        & 1            & 1            & 2 + 2/N = 3                                                                           \\ \hline
3                        & 3            & 3           & 3        & 0            & 0            & 3 + 0 = 3                                                                             \\ \hline
\end{tabular}
\label{Table:2}
\end{table}
\section{Results}
\label{sec:stats}
Here we have presented the comparative analysis of Scopus, WoS, and conflate at author's, organization,  and journal level.
\subsection{At author's level}

\begin{figure}[h!]
    \centering
    \includegraphics[width=0.95\linewidth]{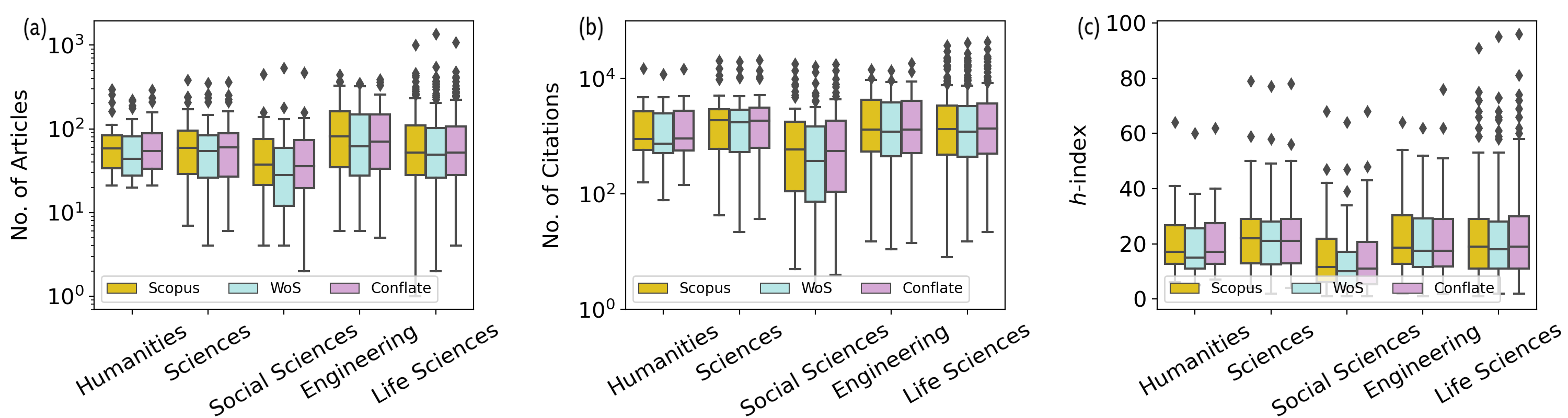}

    \caption{A comparative analysis between Scopus (colored in golden), WoS (colored in cyan), and conflate (colored in pink) for 400 authors as (a) the number of articles, (b) the number of citations,  and (c) $h$-index. The analysis is performed for five disciplines: Humanities, Sciences, Social Sciences, Engineering, and Life Sciences. The standard deviation recorded for Scopus: articles (94.13), citations (4324.97) and $h$-index (14.82); for WoS: articles (105.43), citations (4263.24) and $h$-index (14.56); and for conflate: articles (94.06), citations (4599.34) and $h$-index (15.16).}
    \label{fig2}
\end{figure}

Fig.~\ref{fig2} shows the comparison of conflate with Scopus and WoS on the basis of the number of articles,  the number of citations, and $h$-index of 400 authors. Authors have been categorized into five disciplines (number of authors) based on their work domain: \textit{Social Sciences} (66), \textit{Sciences} (43), \textit{Humanities} (20), \textit{Life Sciences} (211), and \textit{Engineering} (60).  Scopus contains the large number of articles for Social Sciences, Sciences, \textit{Humanities}, and \textit{Engineering} whereas WoS shows for \textit{Life Sciences}.  The number of articles in conflate ranges in between Scopus and WoS count,  except \textit{Life Sciences}.
A large number of citations are reported in \textit{Life Sciences  }in Scopus and the less number of citations are reported in \textit{Social Sciences }in WoS. For \textit{Sciences}, \textit{Engineering}, and \textit{Life Sciences}, conflate has reported the highest number of citations as compared to Scopus and WoS. For the rest of the disciplines, the number of citations reported by conflate is in between the range of Scopus and WoS. For the $h$-index of 400 authors,  we found that conflate has reported the same $h$-index in \textit{Social Sciences} and \textit{Sciences} as reported by Scopus. For \textit{Humanities} and \textit{Engineering}, conflate has reported $h$-index in the range of Scopus and WoS. For \textit{Life Sciences,} Scopus and WoS have reported the same $h$-index whereas conflate has reported one point higher of both.

\begin{figure}[h!]
    \centering
    \includegraphics[width=0.95\linewidth]{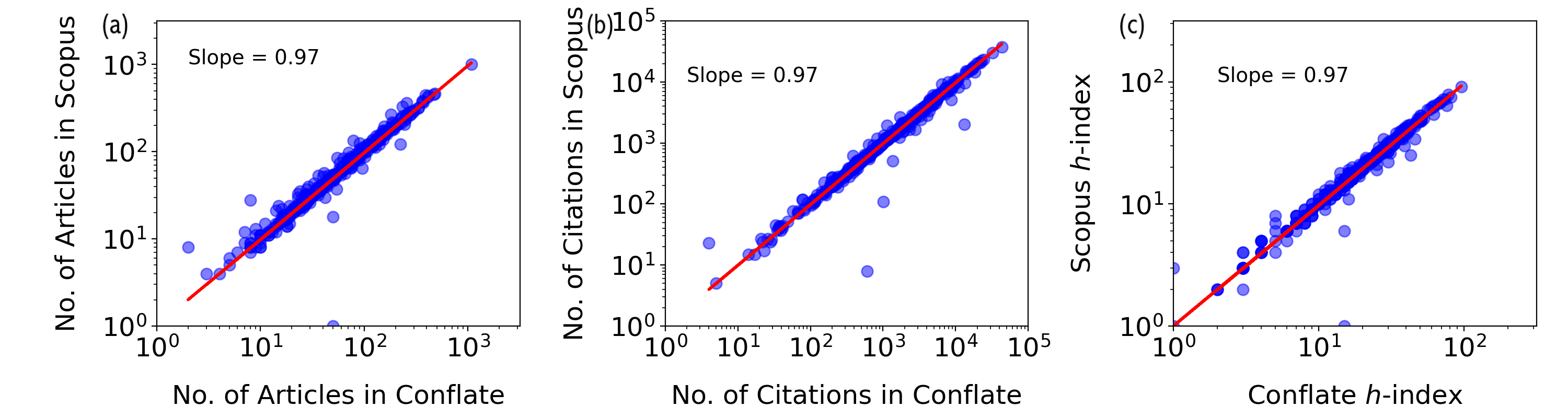}
   \includegraphics[width=0.95\linewidth]{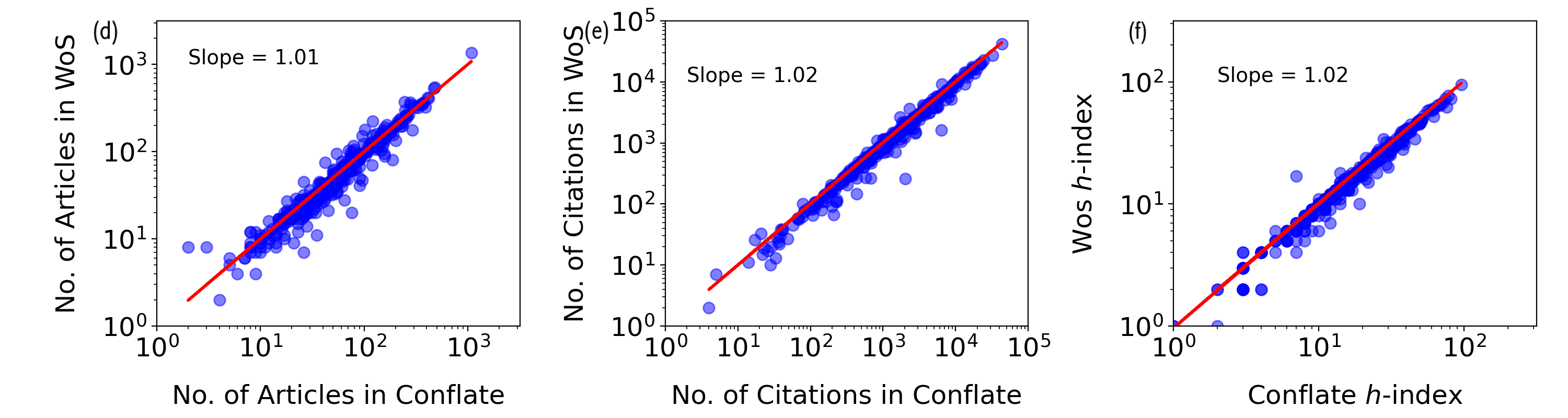}
  \caption{Comparative analysis of 400 authors between Scopus and conflate (a-c) and WoS and conflate (d-f).  The number of articles and citations in Scopus (33182, 1097446), WoS (31732, 1024808), and conflate (32376, 1130306). The maximum $h$-index in Scopus (91), WoS (95 ), and conflate (96). The red line represents the best fit line.}        
    \label{fig3}
\end{figure}
Fig.~\ref{fig3} shows the comparative analysis of the number of articles, citations, and $h$-index between Scopus, WoS, and conflate. The best fit line (colored in red) shows less variation among the Scopus and conflate,   and WoS and conflate. The overall slope is higher in WoS.
In the comparative analysis of Scopus and conflate (see  Fig.~\ref{fig3} (a-c)),  it is observed that the average number of articles published by an author is 83, whereas in conflating it is 81. In Scopus, the average number of citations an author received is 2744 whereas in conflate it is 2826. The average $h$-index of an author in Scopus is 21  and 22 in conflate.  Although the average number of articles calculated in Scopus is less than conflate; however,  an average number of citations and average $h$-index are higher in conflate.  Similarly,  in the comparative analysis of WoS and conflate (see  Fig.~\ref{fig3} (d-f)),  the average number of articles published is 79 as compared to the average of 81 articles per author in conflate. The average number of citations published in WoS is 2562 as compared to 2826 in conflate.  The average $h$-index per author in WoS is 20 whereas in conflating it is 22. Table.~\ref{table:3} represents the comparative analysis of Scopus, WoS, and conflate for 400 author's articles, citations, and $h$- index among different disciplines.

\begin{table}[h!]
\caption{Comparative analysis of Scopus, WoS, and conflate for 400 author's articles, citations and $h$- index for five disciplines.}
\begin{adjustbox}{max width=\textwidth}
\begin{tabular}{|l|c|c|c|c|c|c|c|c|c|}
\hline
\multicolumn{1}{|c|}{}                                                                                                        & \multicolumn{3}{c|}{\textbf{Articles}}                                                                            & \multicolumn{3}{c|}{\textbf{Citations}}                                                                           & \multicolumn{3}{c|}{\textbf{Average $h$-index}}                                                                     \\ \cline{2-10} 
\multicolumn{1}{|c|}{\multirow{-2}{*}{\textbf{\begin{tabular}[c]{@{}c@{}}Authors - 400 \\ (Disciplines)\end{tabular}}}} & \multicolumn{1}{l|}{\textbf{Scopus}} & \multicolumn{1}{l|}{\textbf{WoS}} & \multicolumn{1}{l|}{\textbf{Conflate}} & \multicolumn{1}{l|}{\textbf{Scopus}} & \multicolumn{1}{l|}{\textbf{WoS}} & \multicolumn{1}{l|}{\textbf{Conflate}} & \multicolumn{1}{l|}{\textbf{Scopus}} & \multicolumn{1}{l|}{\textbf{WoS}} & \multicolumn{1}{l|}{\textbf{Conflate}} \\ \hline

\textbf{Life   Sciences}                                                                                                      & 17793                                & 18257                             & 17951                                  & 647698                               & 631244                            & 680537                                 & 22                                   & 22                                & 23                                     \\ \hline

\textbf{Social   Sciences}                                                                                                    & 3531                                 & 3113                              & 3457                                   & 115904                               & 94195                             & 114158                                 & 15                                   & 13                                & 15                                     \\ \hline
 
\textbf{Engineering}                                                                                                          & 6741                                 & 5658                              & 5940                                   & 161092                               & 138631                            & 162218                                 & 23                                   & 21                                & 22                                     \\ \hline

\textbf{Sciences}                                                                                                             & 3397                                 & 3187                              & 3340                                   & 127009                               & 121752                            & 128423                                 & 24                                   & 23                                & 24                                     \\ \hline

\textbf{Humanities}                                                                                                           & 1720                                 & 1517                              & 1688                                   & 45743                                & 38986                             & 44970                                  & 22                                   & 20                                & 22                                     \\ \hline
\end{tabular}
\end{adjustbox}
\label{table:3}
\end{table}

\subsection{At organization level}

\begin{figure}[h!]
    \centering
    \includegraphics[width=0.95\linewidth]{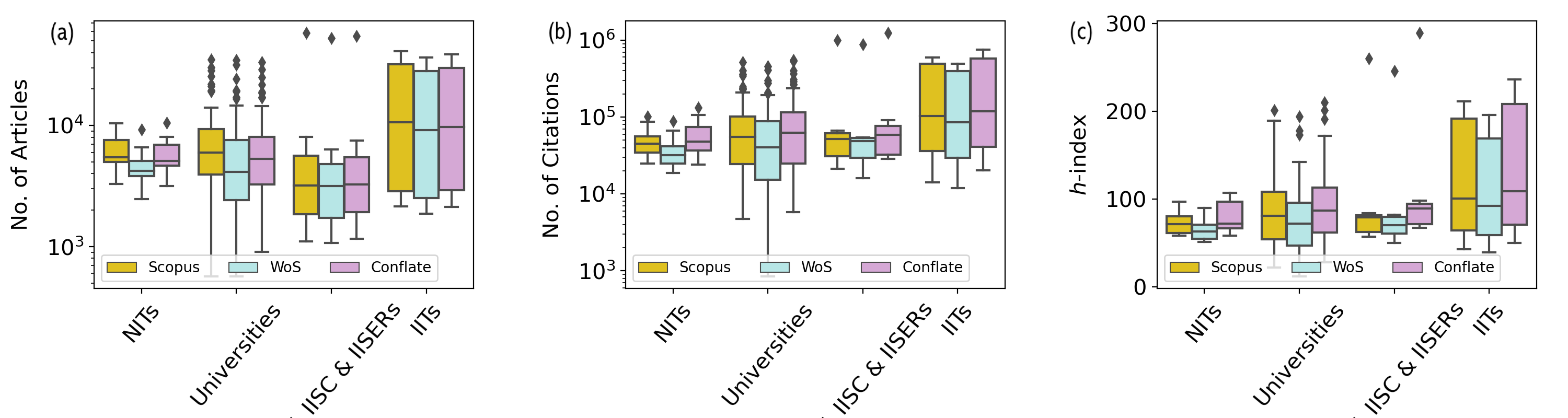}
    \caption{A comparative analysis between Scopus (colored in golden), WoS (colored in cyan), and conflate (colored in pink)for top 100 Indian institutes as (a) the number of articles,  (b) the number of citations,  and (c) $h$-index. The analysis is performed for four categories of institutes: NITs, Universities, IISC \& IISER, and IITs. The standard deviation recorded for Scopus: articles (10459.32), citations (161121.19), and $h$-index (47.36);  for WoS: articles (9637.33), citations (138670.18), and $h$-index (44.24); and for conflate: articles (9774.53), citations (197156.68), and $h$-index (51.75).}
    \label{fig4}
\end{figure}

Here we analyzed the top 100 organizations in India and the categorization is done on the basis of their entity specification (count): Universities (69), IITs (16), NITs (8), and IISC \& IISER (7). It is observed that the highest number of articles published among different databases are from IITs and the lowest number of articles are from NITs. Conflate reported that the number of articles published among different databases is varying between Scopus and WoS across all entities. In all entities, conflate reported the highest number of articles as compared to WoS and the lowest number of articles as compared to Scopus. Conflate reported the highest number of citations as compared to Scopus and WoS for all entities. In comparison with Scopus and WoS, Scopus has always reported more number citations as compared to citations reported by WoS. $h$-index reported by conflate is also highest among both databases. IITs have received the highest $h$-index and NITs have received the lowest $h$-index among other entities. Conflate also reported that the results generated are always in between the range of Scopus and WoS. Among four entities, it can be observed that IITs have the highest $h$-index across multiple databases as shown in Fig.~\ref{fig4}.  Further, one can analyze the different disciplines of these organizations to keep track of the most popular discipline in terms of research publications.

\begin{figure}[h!]
    \centering
    \includegraphics[width=0.95\linewidth]{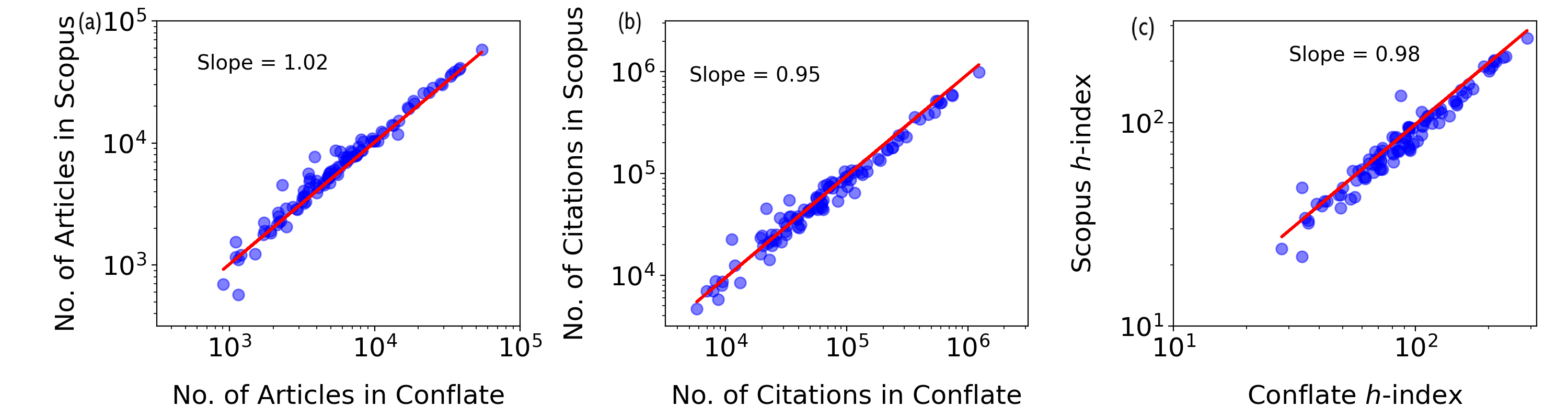}
        \includegraphics[width=0.95\linewidth]{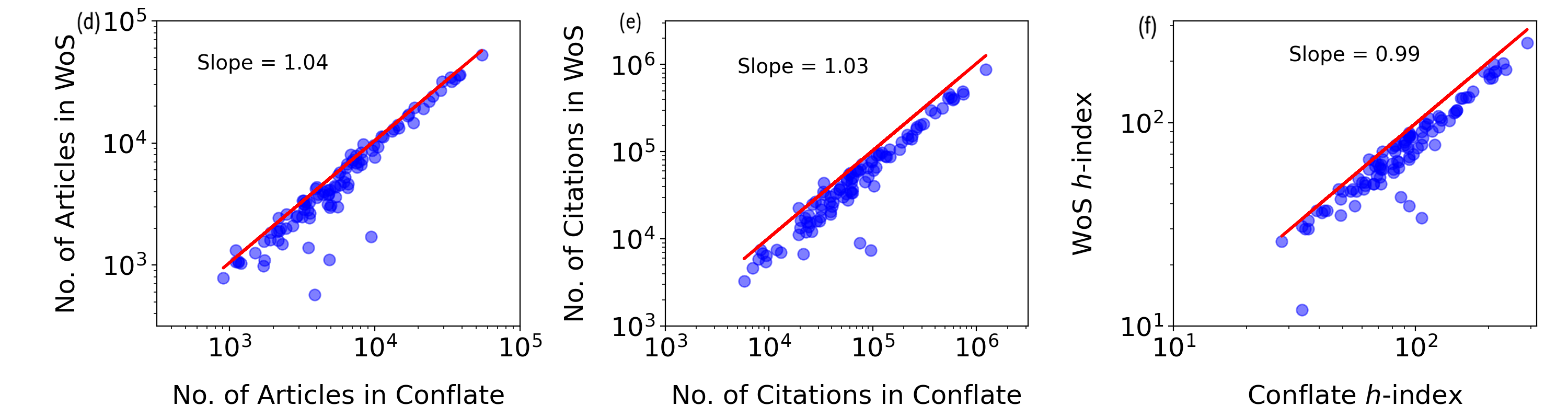}
    \caption{Comparative analysis of top 100 organizations between Scopus and conflate (a-c) and WoS and conflate (d-f).  The number of articles and citations in Scopus (964093, 11399909), WoS (797158,  9337059), and conflate (873719, 13483112). The maximum $h$-index in Scopus (260), WoS (246), and conflate (289). The red line represents the best fit line.}
    \label{fig5}
\end{figure}

Fig.~\ref{fig5} shows the comparative analysis between Scopus, WoS, and conflate for top 100 Indian organizations. 
In the comparative analysis of Scopus and conflate (see Fig.~\ref{fig5} (a-c)),  it is observed that the average number of articles in Scopus is 9641 as compared to 8737 in conflate.  The difference in the average number of articles states that all articles published in Scopus are not considered in conflate. The average number of citations recorded in Scopus is 113999 as compared to 134831 in conflate.  The average $h$-index calculated in Scopus for these organizations is 91 which is quite lesser than the average $h$-index (100) calculated in conflate.  Similarly, in the comparative analysis of WoS and conflate (see  Fig.~\ref{fig5} (d-f)),  it is observed that the average number of articles in WoS is 7971 whereas in conflating it is 8737. Conflate also reported a significantly higher number of citations with an average score of 134831 as compared to 93371 in WoS. The average $h$-index in conflate is also 100 which is quite higher than the average $h$-index 82 reported by WoS. Table.~\ref{table:4} represents the comparative analysis of Scopus, WoS, and conflate for articles, citations, and $h$- index for 100 organizations categorized into 4 main head organizations. 

\begin{table}[!h]
\caption{Comparative analysis of Scopus, WoS, and conflate for articles, citations and $h$- index for 100 organizations categorized in to 4 main head organizations.  }
\begin{adjustbox}{max width=\textwidth}
\begin{tabular}{|l|c|c|c|c|c|c|c|c|c|}
\hline
\multicolumn{1}{|c|}{}                                                                                                      & \multicolumn{3}{c|}{\textbf{Articles}}                                                                            & \multicolumn{3}{c|}{\textbf{Citations}}                                                                           & \multicolumn{3}{c|}{\textbf{Average $h$-index}}                                                                     \\ \cline{2-10} 
\multicolumn{1}{|c|}{\multirow{-2}{*}{\textbf{\begin{tabular}[c]{@{}c@{}}Organizations - 100 \\ (Type Wise)\end{tabular}}}} & \multicolumn{1}{l|}{\textbf{Scopus}} & \multicolumn{1}{l|}{\textbf{WoS}} & \multicolumn{1}{l|}{\textbf{Conflate}} & \multicolumn{1}{l|}{\textbf{Scopus}} & \multicolumn{1}{l|}{\textbf{WoS}} & \multicolumn{1}{l|}{\textbf{Conflate}} & \multicolumn{1}{l|}{\textbf{Scopus}} & \multicolumn{1}{l|}{\textbf{WoS}} & \multicolumn{1}{l|}{\textbf{Conflate}} \\ \hline

\textbf{NITs}                                                                                                               & 50362                                & 39059                             & 47683                                  & 416302                               & 318676                            & 492760                                 & 74                                   & 66                                & 80                                     \\ \hline

\textbf{Universities}                                                                                                       & 565887                               & 451489                            & 499159                                 & 6051897                              & 4917831                           & 6997951                                & 85                                   & 76                                & 93                                     \\ \hline

\textbf{IISC \& IISER}                                                                                             & 77349                                & 70063                             & 73835                                  & 1252987                              & 1107018                           & 1542242                                & 98                                   & 92                                & 111                                    \\ \hline

\textbf{IITs}                                                                                                               & 270495                               & 236547                            & 253042                                 & 3678723                              & 2993534                           & 4450159                                & 122                                  & 110                               & 134                                    \\ \hline
\end{tabular}
\end{adjustbox}
\label{table:4}
\end{table}

\subsection{Journal level}

\begin{figure}[h!]
    \centering
    \includegraphics[width=0.95\linewidth]{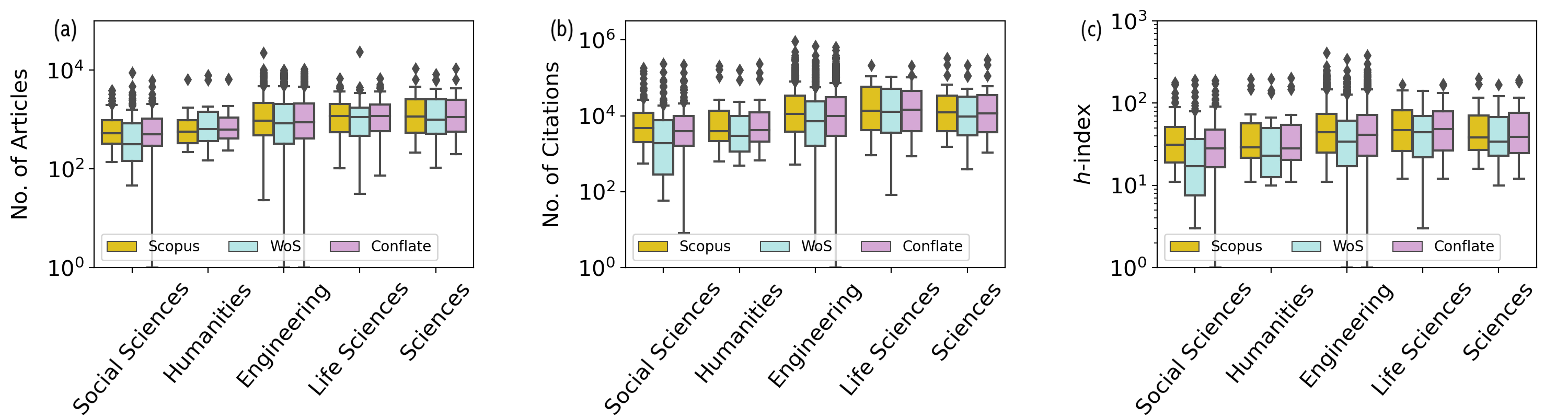}
    \caption{A comparative analysis between Scopus (colored in golden), WoS (colored in cyan), and conflate (colored in pink) of 1000 journals as (a) the number of articles,  (b) the number of citations,  and (c) $h$-index. The analysis is performed for five disciplines: Humanities, Sciences, Social Sciences, Engineering, and Life Sciences. The standard deviation recorded for Scopus: articles (1794.63), citations (59090.78), and $h$-index (44.40);  for WoS : articles (1819.10), citations (45753.64), and $h$-index (40.06); and for conflate: articles (1677.79), citations (55723.79), and $h$-index (45.13).  }
    \label{fig6}
\end{figure}

Here we analyzed 1000 journals and broadly divided into 5 disciplines (journal count), \textit{Engineering} (800), \textit{Social Sciences} (119), \textit{Life Sciences} (35), \textit{Sciences} (27), and \textit{Humanities} (19). The number of articles observed in \textit{Sciences} is highest in Scopus with lowest in \textit{Social Sciences}. For \textit{Social Sciences}, conflate reported the highest number of articles among Scopus and WoS. For \textit{Humanities}, \textit{Engineering} and \textit{Sciences}, conflate has reported a number of articles in between Scopus and WoS. For \textit{Life Sciences}, conflate has reported almost the same number of articles as compared to Scopus which is quite lesser than the WoS database. The number of citations reported by conflate is in between the range of Scopus and WoS for all disciplines where \textit{sciences} is on top and \textit{social sciences} is at the bottom. $h$-index reported by conflate for 1000 journals is the same as reported by Scopus for \textit{Humanities}, \textit{Sciences}, and \textit{Life Sciences}. For \textit{Social Sciences} and \textit{Engineering}, it is in between the range of Scopus and WoS. Lowest $h$-index is reported by WoS for \textit{Social Sciences} and highest by Scopus for \textit{Life Sciences} as shown in Fig.~\ref{fig6}.
\begin{figure}[h!]
    \centering
    \includegraphics[width=0.95\linewidth]{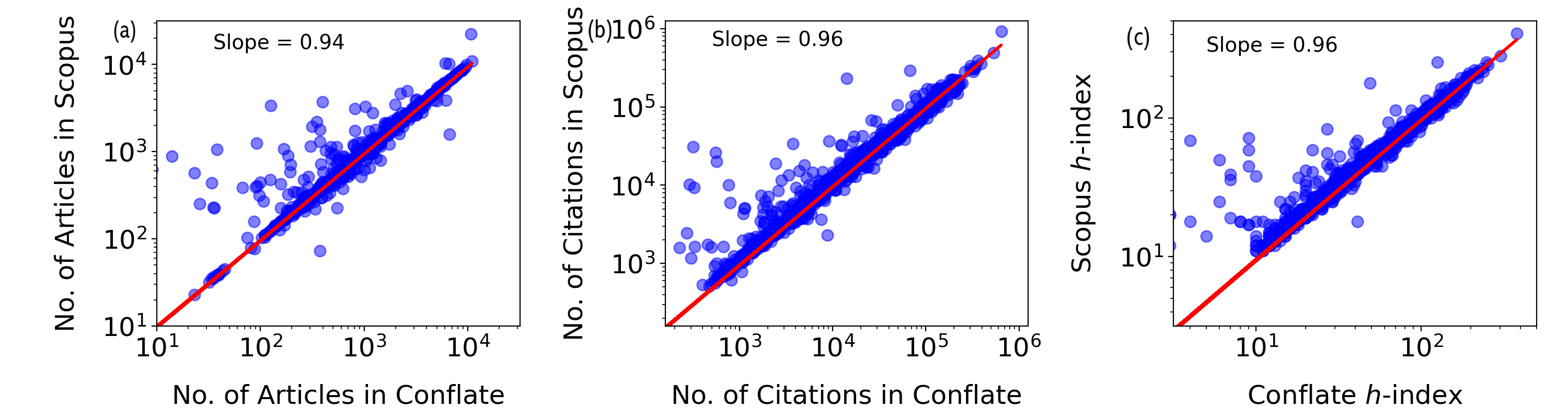}
      \includegraphics[width=0.95\linewidth]{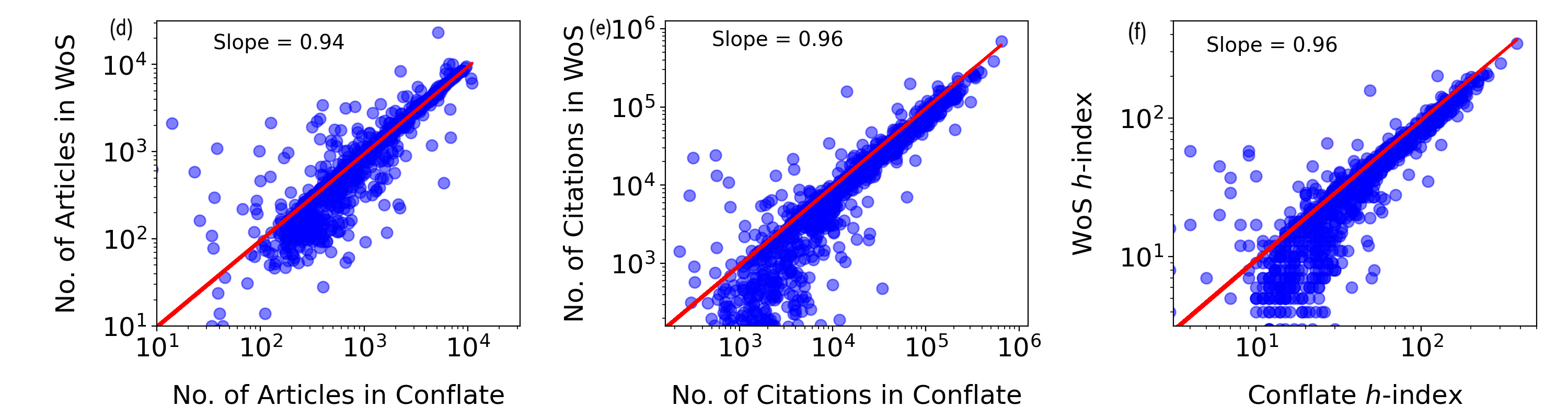}
    \caption{Comparative analysis of 1000 journals between Scopus and conflate (a-c) and WoS and conflate (d-f).  The number of articles and citations in Scopus (1528904, 30964292), WoS (1415093, 22570461), and conflate (1481823, 29276118). The maximum $h$-index in Scopus (408), WoS (344 ), and conflate (381). The red line represents the best fit line.}
    \label{fig7}
\end{figure}
Fig.~\ref{fig7} shows the comparative analysis between Scopus, WoS, and conflate for  1000 journals. 
In the comparative analysis of Scopus and conflate (see Fig.~\ref{fig7} (a-c)),  it is observed that the average number of articles in Scopus is 1529 and in conflate is 1482.  There is a slight hike in the average number of articles in Scopus.  Similarly, the average number of citations in Scopus is 30964, and in conflate is 29276.  The average $h$-index  calculated in Scopus is 56 and in conflate is 53.  Similarly,  in the comparative analysis of WoS and conflate (see Fig.~\ref{fig7} (d-f)),  it is observed that the average number of articles in WoS (1415) as compared to conflate (1482) shows significantly close values. The average number of citations in WoS is 22570 as compared to 29276 of conflate. Conflate clearly states that there is more scope of consideration of citations as compared to citations considered by WoS. The average $h$-index in WoS is 44 as compared to 53 in conflate. Conflate is clearly moving ahead in terms of the average number of citations and $h$-index calculation of WoS.
Table.~\ref{table:5} represents the comparative analysis of Scopus, WoS, and conflate for 1000 journal articles, citations, and $h$- index among different disciplines.


\begin{table}[!h]
\caption{Comparative analysis of Scopus, WoS, and conflate for 1000 journals articles, citations and $h$- index for different disciplines.}
\begin{adjustbox}{max width=\textwidth}
\begin{tabular}{|l|c|c|c|c|c|c|c|c|c|}
\hline
\multicolumn{1}{|c|}{}                                                                                                          & \multicolumn{3}{c|}{\textbf{Articles}}                                                                            & \multicolumn{3}{c|}{\textbf{Citations}}                                                                           & \multicolumn{3}{c|}{\textbf{Average $h$-index}}                                                                     \\ \cline{2-10} 
\multicolumn{1}{|c|}{\multirow{-2}{*}{\textbf{\begin{tabular}[c]{@{}c@{}}Journals - 1000 \\ (Disciplines)\end{tabular}}}} & \multicolumn{1}{l|}{\textbf{Scopus}} & \multicolumn{1}{l|}{\textbf{WoS}} & \multicolumn{1}{l|}{\textbf{Conflate}} & \multicolumn{1}{l|}{\textbf{Scopus}} & \multicolumn{1}{l|}{\textbf{WoS}} & \multicolumn{1}{l|}{\textbf{Conflate}} & \multicolumn{1}{l|}{\textbf{Scopus}} & \multicolumn{1}{l|}{\textbf{WoS}} & \multicolumn{1}{l|}{\textbf{Conflate}} \\ \hline

\textbf{Life   Sciences}                                                                                                        & 60061                                & 68142                             & 60049                                  & 1144780                              & 858220                            & 1130871                                & 59                                   & 50                                & 59                                     \\ \hline

\textbf{Social   Sciences}                                                                                                      & 94755                                & 86719                             & 98316                                  & 1712570                              & 1268836                           & 1569405                                & 42                                   & 30                                & 38                                     \\ \hline

\textbf{Engineering}                                                                                                            & 1298929                              & 1179771                           & 1244492                                & 26468041                             & 19176940                          & 24982597                               & 58                                   & 46                                & 55                                     \\ \hline

\textbf{Sciences}                                                                                                               & 56390                                & 53458                             & 54009                                  & 1061160                              & 783108                            & 1030698                                & 56                                   & 49                                & 56                                     \\ \hline

\textbf{Humanities}                                                                                                             & 18769                                & 27003                             & 24957                                  & 577741                               & 483357                            & 562547                                 & 54                                   & 46                                & 53                                     \\ \hline
\end{tabular}
\end{adjustbox}
\label{table:5}
\end{table}
\section{Technological Enhancement}
\label{sec:tech}

Distributed ledger technology has found its applications in the field of education for verification of academic records \citep{aamir2020blockchain},  sharing of student credentials \citep{mishra2020implementation}, adoption of smart learning environments \citep{ullah2021blockchain} and in implementation of mobile-based higher education systems \citep{arndt2020blockchain}.
This gives us an opportunity to work on the potential of this technology and enhance its features to its full capability. in the field of education.  Making the use of distributed ledger technology in the research publication industry is a novel approach. There is a very good scope of observing challenges associated with new implementations \citep{upadhyay2020demystifying}. Features like decentralization, persistency, anonymity, and auditability of records give more confidence to its stakeholders in a system presenting a scientific work of authors, organizations, and journals \citep{mclean2016demystifying, deshpande2017distributed}. Hence, using DLT in the research publication industry can be considered as a viable choice to systematically achieve a sustained system in the interest of its stakeholders. To achieve acceptability in DLT the consensus is used where every individual in a system must accept the happenings \citep{bamakan2020survey}. There are multiple studies on the performance and analysis of various consensus algorithms \citep{mingxiao2017review, hao2018performance, yim2018blockchain, bach2018comparative}. To achieve consensus in unified informetrics, a concept of summation of common and unique citations may be used.

\subsection{Mapping of publications and DLT}

\begin{figure}[h]
   \centering
   \includegraphics[width=0.95\linewidth]{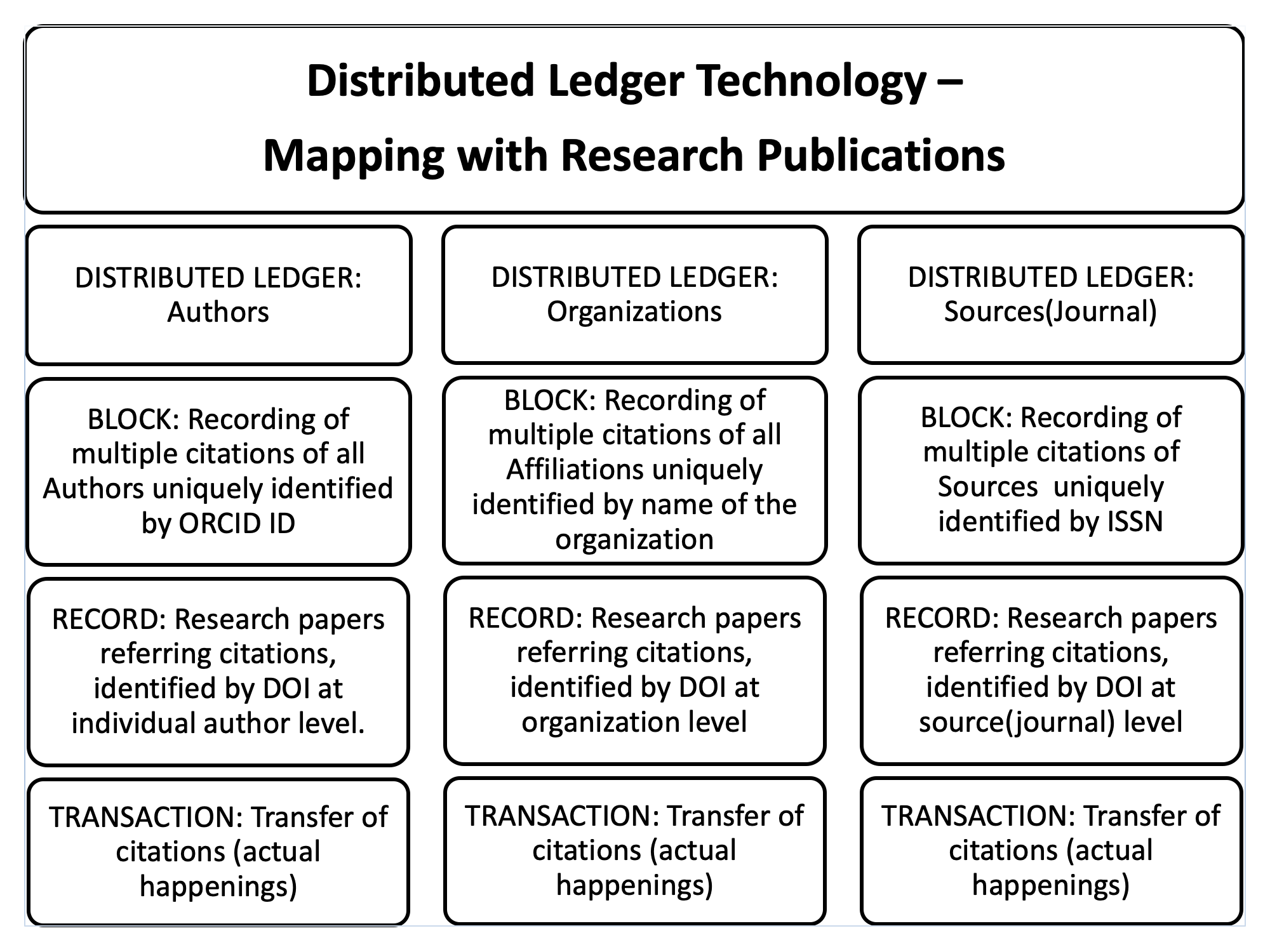}
   \caption{Schematic representation of DLT mapping with research publications. An informetric ledger will contain the individual blocks for all the entities like authors, affiliations, and journals. A conflate ledger block will be created on the basis of articles referred to as records and citations referred to as a transaction for each individual entity. Block can be considered as the conflate of records and transactions whereas a ledger can be considered as the conflate of block, record, and transactions.
}
   \label{fig8}
\end{figure}
A distributed ledger data bank is believed and accessed autonomously by every contributor involved in a huge system \citep{maull2017distributed}. The allotment is exclusive and records are held and autonomously constructed by each node \citep{collomb2016blockchain}. A block acts as a page of a register or record book. It is thus a stable collection of records which, once engraved, cannot be reformed or removed \citep{sunyaev2020distributed}. The blocks are supplemented to the chain in a direct-sequential order. Chain structure permanently time-stamp's and stores exchange of value, preventing anyone from altering the ledger~\citep{naughton2016blockchain, rauchs2018distributed}. Each block's record indicates a minimum of one transaction; however, many effective transactions can be characterized in a single block. Every transaction record (ledger entry) is connected to preceding transactions and is consistent for every contributing node \citep{olnes2017blockchain}. Every ledger entry is retraceable through its complete antiquity and can be remodeled \citep{liu2020distributed}. Fig.~\ref{fig8} gives the insight details that how the terminologies in DLT may be mapped with academic research publications. To map DLT with academic research publications, we propose three ledgers as follows:
\begin{itemize}
   \item \textbf{Author ledger:} In the author ledger, research contributions identified with DOI will act as a record, and citations received by such contributions will act as a Block. Transfer of citations in the author ledger will be considered as transactions. The uniqueness of the author ledger will be maintained by the ORCID ID of the authors.
           \item \textbf{Organization ledger:} In the organization ledger, research contributions identified with DOI will act as a record, and citations received by such contributions will act as a Block. Transfer of citations in the organization ledger will be considered as transactions. The uniqueness of the organization ledger will be maintained by the organization name.
           \item \textbf{Journal ledger:} In the journal ledger, research contributions identified with DOI will act as a record, and citations received by such contributions will act as a Block. Transfer of citations in the journal ledger will be considered as transactions. The uniqueness of the journal ledger will be maintained by journal ISSN.
\end{itemize}
\subsection{Implementation process}
We have implemented the distributed ledger in Python.  It is the fully functional distributed ledger-based application~\citep{ibm_blockchain_post}. It uses the concept of Python, HTML, and Javascript to build up the required framework where an individual can interact with the system through its web browser. The steps of implementation are:
\begin{enumerate}
\item Selection of profile as an author, organization, or journal.
\item Input ORCID or Organization name or ISSN.
\item Fetch the required publication and citation details from both bibliographic databases.
\item Do citation analysis.
\item Generate single publication and citation count of the selected profile.
\item Start DLT server on the available port, usually on 8000.
\item Start DLT based application on the available port, usually on 5000.
\item Browse the csv file generated for single publication and citation count of the selected profile.
\item Click on the Post button to post the request to the DLT server.
\item Click on the button ``Request to mine'' to mine the node in the ledger.
\item Click on the button ``Resync'' to resync with the blocks for updated data on the ledger.
\item Blocks will be displayed with the calculated $h$-index on the basis of single publication and citation count uploaded for the selected profile.
\end{enumerate}

\subsection{Implications of the proposed approach}
In the current scenario, different bibliographic databases like Scopus,  WoS,  etc. have their own systems where an author and organization can find their scientific impact \citep{visser2021large}. This makes the process quite tedious, as one ends up with three different citation counts and three different $h$-index values for the same entities due to an individual calculation system of these bibliographic databases. The proposed system provides a single platform that fetches the required information of publications and citations from multiple bibliographic databases, performs analysis on the fetched data, and provides unified informetrics. The proposed system provides:
\begin{enumerate}
\item Cognitive and synthesized informetrics.
\item Expands the knowledge base of its stakeholders by comprising the information of multiple bibliographic databases.
\item Explores informetrics in a novel way and gives a clear assessment of research impact.
\item Supports the integration of $N$ bibliographic databases.
\item Presents in-depth analyses of the core components like publications, citations, self-citations,  etc.
\item Supports better understanding of research impact of authors, organizations,  and journals.
\item Facilitates its stakeholders for the establishment of a system providing a clear, authentic, and simulated environment for the research measurement of entities.
\end{enumerate}
\section{Summary}	
\label{sec:conclusion}
Scopus and WoS are considered as an important database which is being used worldwide. Both are traditional and the authentic source of accessing scientific work \citep{zhu2020tale}. Different universities, government organizations, recruiters, accreditation, and ranking agencies ask informetrics based on Scopus and WoS separately in their job applications and documentation.  Authors are required to provide total publication count, citation count,  $h$-index, etc. both from Scopus and WoS separately. This gives different informetrics for a single author.  There is no common platform in our knowledge that can record or calculate single informetrics across multiple bibliographic databases. Therefore, a weighted unified informetrics system based on distributed ledger technology named ``conflate'' has been discussed and proposed. The proposed solution provides a transparent and distributed view of the research contributors to their stakeholders. Calculated results also signify the efficiency of ``conflate''. We have used Scopus and WoS for the implementation due to the availability of the data. The proposed implementation of the pay-off matrix also strengthens the overall framework of the proposed system.

The key findings of the work are: (i) It presents a unified method to maintain records associated with entities of author, organization, and journal. This method determines an absolute number of articles and citations for different entities. (ii) The mapping of multiple bibliographic databases for the calculation of $h$-index, and related informetric with the concept of pay-off matrix.  (iii) The use of distributed ledger technology for the generation of ledger blocks for authors, organizations, and journals to safeguard their research contributions for any kind of manipulation and provides a robust platform for the presentation.

The presented work has some advantages as (i) The DOI-based data filtration helps us to identify the authenticity of received citations and publications~\citep{gorraiz2016availability}.  (ii) Different stakeholders like government agencies, accreditation agencies, ranking organizations, and funding agencies can use the proposed system for the evaluation of the research contribution of individuals, organizations as well as journals.  (iii)The introduction of distributed ledger technology in research publications will provide a verifiable research record of entities across different ledgers. This will introduce a tempering proof system for the integrity of records.  (iv) The proposed system is a novel system introduced with the conflate of two traditional bibliographic databases like Scopus and WoS. 

The major limitation of the study is the fact that we have considered the publications where DOI exists. In case WoS and Scopus do not have DOI numbers for the particular publications, we will not be able to consider the publication as authentic and the author will lose publications count and their citations count as well. Moreover, it could be a citation loss for low profile authors who have their work indexed only in Scopus or in WoS, refer to example 1 in Table.~\ref{Table:2}. For such journals which are indexed only in Scopus or in WoS, also shows their limitations to other bibliographic databases. If an author publishes his work in a journal that is indexed in multiple bibliographic databases, there is a good chance of higher visibility of a scientific work to be read and cited worldwide.  As new bibliographic databases may populate in the near future, the proposed system should support the integration of those databases into the existing system.

To conclude further, there are still several possible areas for further exploration and extension. Here are some interesting areas for possible future developments and research.
\begin{itemize}
	\item \textit{Different bibliographic databases:}
	We have studied the features of two bibliographic databases such as Scopus and WoS. Hence, the performed study is limited to two bibliographic databases. One can extend the study further with the use of bibliographic databases like Microsoft Academics, Google Scholar \citep{martin2021google}, OpenAIRE \citep{rettberg2015openaire}, DataCite \citep{brase2009datacite}, Mendeley \citep{reiswig2010mendeley} and Zenodo \citep{peters2017zenodo}. All these bibliographic databases may create conflate as per the model to calculate unified informetrics.\\
   \item \textit{Different technological aspects:}
   We have empowered citation analysis with distributed ledger technology. Hence, the performed study is limited to the implementation of only one technology in the research publishing industry. One can extend the study further with the use of ``Gamification'' and its gaming elements in the research publishing industry. The use of Gamification in the research publication industry can be helpful to increase the motivation and encouragement of its stakeholders for the extraction of unified informetrics for different entities \citep{kumar2012gamification}.
   
\end{itemize}

\begin{acknowledgements}
We would like to thank Dr. Ranbir Singh Batth and Mr. Sukhvir Singh for help and support. Both Scopus and WoS data has been downloaded from the Northwestern University, USA.
\end{acknowledgements}
%
\section*{Conflict of interest}
The authors declare that they have no conflict of interest.

\bibliographystyle{spbasic}      
\bibliography{cas-refs}   

%
%

\end{document}